\documentclass[sigplan,screen,balance=false,authorversion=true]{acmart}

\AtBeginDocument{%
  \providecommand\BibTeX{{%
    \normalfont B\kern-0.5em{\scshape i\kern-0.25em b}\kern-0.8em\TeX}}}

\usepackage{fancyvrb}
\usepackage{graphicx}

\setcopyright{rightsretained}
\acmYear{2023}\copyrightyear{2023}
\acmConference[CF '23]{20th ACM International Conference on Computing Frontiers}{May 9--11, 2023}{Bologna, Italy}
\acmBooktitle{20th ACM International Conference on Computing Frontiers (CF '23), May 9--11, 2023, Bologna, Italy}
\acmDOI{10.1145/3587135.3592191}
\acmISBN{979-8-4007-0140-5/23/05}

\begin{document}

\title{QHDL: a Low-Level Circuit Description Language for Quantum Computing}

\author{Gilbert Netzer}
\orcid{0000-0002-9479-7393}
\affiliation{%
  \institution{KTH Royal Institute of Technology}
  \city{Stockholm}
  \country{Sweden}}
\email{noname@kth.se}

\author{Stefano Markidis}
\orcid{0000-0003-0639-0639}
\affiliation{%
  \institution{KTH Royal Institute of Technology}
  \city{Stockholm}
  \country{Sweden} }
\email{markidis@kth.se}

\renewcommand{\shortauthors}{Netzer and Markidis}

\begin{abstract}
This paper proposes a descriptive language called QHDL, akin to VHDL, to program gate-based quantum computing systems. Unlike other popular quantum programming languages, QHDL targets low-level quantum computing programming and aims to provide a common framework for programming FPGAs and gate-based quantum computing systems. The paper presents an initial implementation and design principles of the QHDL framework, including a compiler and quantum computer simulator. We discuss the challenges of low-level integration of streaming models and quantum computing for programming FPGAs and gate-based quantum computing systems.
\end{abstract}



\maketitle

\section{Introduction}

Quantum computing is a potentially disruptive computing paradigm that exploits
the quantum mechanical behavior of atom scale systems to provide a
\emph{quantum advantange} over classical large-scale computations.
Prominent applications include cryptology, e.g. Shor's algorithm to factor
integer numbers~\cite{gidney2021factor}, search in unstructured data, e.g. Grover's algorithm, and linear algebra, e.g. the HHL algorithm. Hybrid algorithms, such as Shor's algorithm or the quantum variational eigensolvers~\cite{cerezo2021variational} or emerging quantum machine learning~\cite{schuld2018supervised},
partition the computation into both a quantum and classical part that work together to solve a given problem.
\begin{figure}[b]
  \includegraphics[width=70mm]{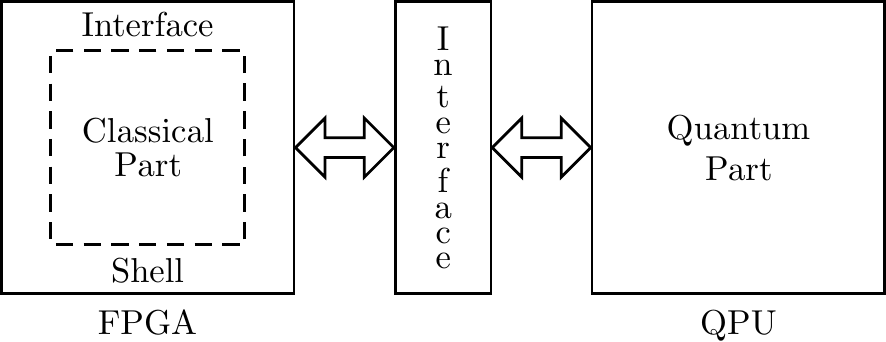}
  \caption{\label{fig:system}Block-diagram of a tightly integrated quantum computing system.}
  \Description{A block-diagram of three major blocks arranged from left to right. The leftmost block, labelled FPGA, contains the classical part of the algorithm surrounded by an interface shell. It bidirectionally communicates with the middle block labelled interface. In turn the middle block communicates bidirectionally with the rightmost block labelled QPU. This block contains the quantum part of the algorithm.}
\end{figure}
In contrast to today’s loosely integrated quantum computing
systems, where the classical processing is carried out on a
general-purpose computer, we envision that tightly integrated systems, as shown in Figure~\ref{fig:system}, where at least some parts of the classical processing take place in dedicated hardware close to the quantum processor become more viable. 
Such systems could, for instance, be helpful to process real-time or streaming data with low latency and high bandwidth requirements, to carry out more advanced statistic gathering from repeated quantum computations, or also to implement closed-loop hybrid algorithms where the quantum computation is partially observed and control fed back to the Quantum Processing Unit (QPU) while a quantum state in superposition has to be maintained in the quantum algorithm part.
Suitable hardware has already been constructed by others~\cite{xu2021a}, although with a different focus.

In this work we specifically focus on a highly optimized integration between the classical and quantum parts, in contrast to popular approaches such as Qiskit or Cirq~\cite{heim2020quantum}. We use an implicit formulation of the sequence in which to carry out the elementary operations as opposed to explicit ISA approaches, such as QASM~\cite{cross2017open} and Quil~\cite{smith2016practical}.
While existing frameworks like Quingo~\cite{fu2021a} can handle real-time integration of quantum computing with instruction-based classical computing, we propose to utilize the fine-grained timing capabilities of hardware description languages used for gate or register-transfer level modelling.

The main contributions of this work are the definition of the QHDL language in Section~\ref{sec:qhdl} and the implementation of a proof-of-concept QHDL/VHDL co-simulation environment in Section~\ref{sec:cosim}. We provide an example of the application of QHDL in Section~\ref{sec:bell}.

\section{Background}

The quantum circuit model of computation is perhaps the most established abstraction of quantum computing. In this model complex quantum computations are composed from elementary quantum gates which are connected to form circuits similar to digital logic circuits. Examples of quantum gates include single-qubit NOT and Hadamard gates, multi-qubit gates such as the controlled-NOT, Toffoli and Fredkin gates, and measurement and preparation operations to interact with classical computations. The connectivity of the quantum circuit also defines the sequence in which the elementary operations have to be carried out and which quantum bits, or qubits, are to be manipulated. Quantum circuits can also be hierarchically composed to form complex algorithms based on simpler primitives such as quantum arithmetic, Amplitude Amplification (AA), or Quantum Fourier Transform (QFT) circuits, which could also be parameterized for instance in the number of qubits used.

\begin{figure}[t]
  \includegraphics[width=70mm]{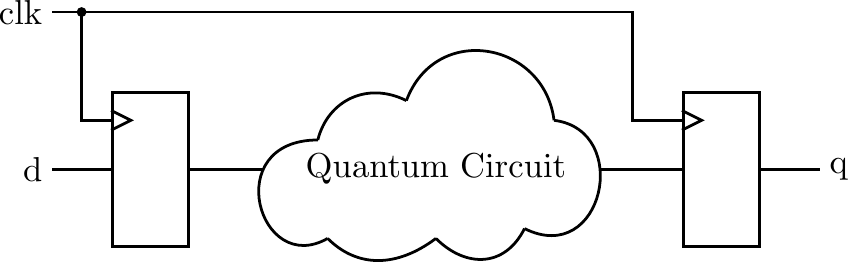}\\[1.5ex]
  \includegraphics[width=70mm]{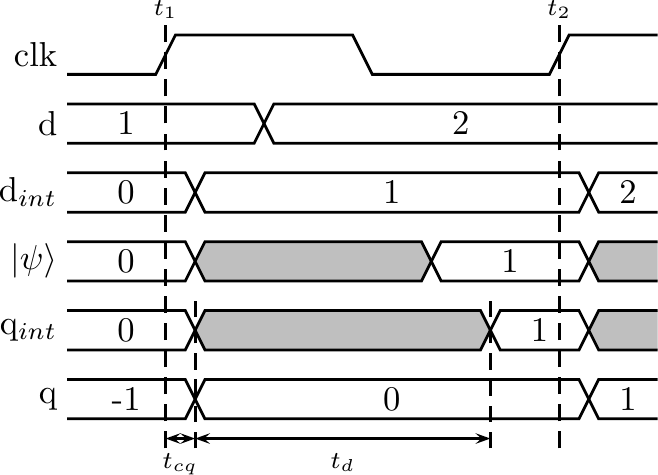}
  \caption{\label{fig:timingmodel}Model of the timing of the in QHDL described circuit in relation to its VHDL environment.}
  \Description{On top a circuit diagram with an external input, labelled d, connected to the input of a register on the left. The output of this register is connected to a logic cloud, labelled quantum circuit, and the output from this could is connected to the input of another register on the right. The output from this register is connected to the external output, labelled q. Both registers are connected to a common clock input, labelled clk. Below a timing diagram for the shown circuit. The topmost signal,
  labelled clk, has two rising edges, labelled t1 and t2. The next signal below, the external input labelled d, changes from state 1 to state two during the high period of clk after t1. The next signal, the output from the input register labelled dint, changes from state 0 to 1 and finally to 2 after delay tcq after t1 respective t2. The quantum state of the simulation represented by the ket-vector psi is shown next, it changes from state 0 to unknown at tcq after t1 and to state 1 before t2. The output from the quantum circuit, labelled qint, also change from state 0 to unknown at tcq after t1 and to state 1 at td after t1, which is before
  t2. The external output labelled q shown last changes from state -1 to 0 and finally 1 at tcq after t1 respective t2.}
\end{figure}

Hardware Description Languages (HDLs) such as Verilog and
VHDL~\cite{ieee1076-2019} were created to support simulation of
complex digital electronics systems by means of concurrently
executing processes to abstract behavior that interact using signals
to abstract electrical wires.
Changes to a signals value, called events, are associated with discrete points
in time to synchronize the constituent processes forming a discrete event
simulation model.
Suitably restricted HDL models or programs can also be used to synthesize
hardware, both in the form of programmable logic such as FPGAs and hardwired
ASIC implementations.
Co-simulation approaches, where independent simulations at different levels
of detail are coupled, e.g.~\cite{zivojnovic1996a}, are also common and
standardized interfaces such as VHPI for VHDL exist.

\section{The QHDL Language}
\label{sec:qhdl}

The QHDL language is used to describe a quantum circuit at the gate level using
the syntax and most of the semantics of a subset of the 2018 version of the
VHDL language~\cite{ieee1076-2019}:
\textit{i)} Entities and associated architecture bodies are used to
describe circuits.
\textit{ii)} Component instantiation is used incorporate built-in quantum gates
or user-defined sub-circuits.
\textit{iii)} Signals are used to connect outputs and inputs of gates to each
other or to ports describing external connections.

To convey the specific semantics of quantum circuits, a new fundamental type
called \texttt{qbit} is defined that models a quantum bit, or qubit.
QHDL also defines three semantic rules for quantum circuits:
\textit{i)} To enforce the ``no-cloning'' theorem of quantum computing,
\texttt{qbit} signals must be driven by a single output and connect to a
single input.
\textit{ii)} The top-level entity that interfaces with the VHDL simulation
must not have ports of the \texttt{qbit} type.
\textit{iii)}  QHDL entities cannot contain any non-quantum logic in the form
of process or concurrent statements or of VHDL components.

Clock inputs are provided on both setup and measure components to allow to
synchronize the exchange of information between the quantum and classical
circuits.
As shown in Figure~\ref{fig:timingmodel} this conceptually encloses the
quantum circuit between two sets of registers providing a timing
independent of the details of the quantum computation:
At each rising edge of the external clock signal, for example at $t_1$,
the inputs, $\mathrm{d}$, provided by the VHDL simulation are sampled and
after $t_{cq}$ applied to the quantum circuit, via $\mathrm{d}_{int}$,
resulting in updates to the state vector, $\mathrm{| \psi \rangle}$, and the
internal measurement results, $\mathrm{q}_{int}$.
After a delay, $t_d$, the computation is completed and the updated
results are sampled at the next clock cycle, $t_2$, and sent back to the
VHDL simulation via the outputs, $\mathrm{q}$.
If gates have no associated delay, as in the current implementation, an
infinitesimal delay is added, similar to a \emph{delta cycle} in VHDL,
to ensure that the updated output always will be presented to the VHDL
simulation at the next clock cycle.

QHDL also defines a standard library, called \texttt{qhdl}, that provides
built-in types and quantum gates, including setup and measure operations that
interact with external classical inputs, in a package called \texttt{std},
similar to VHDL.

\begin{figure}
  \includegraphics[width=70mm]{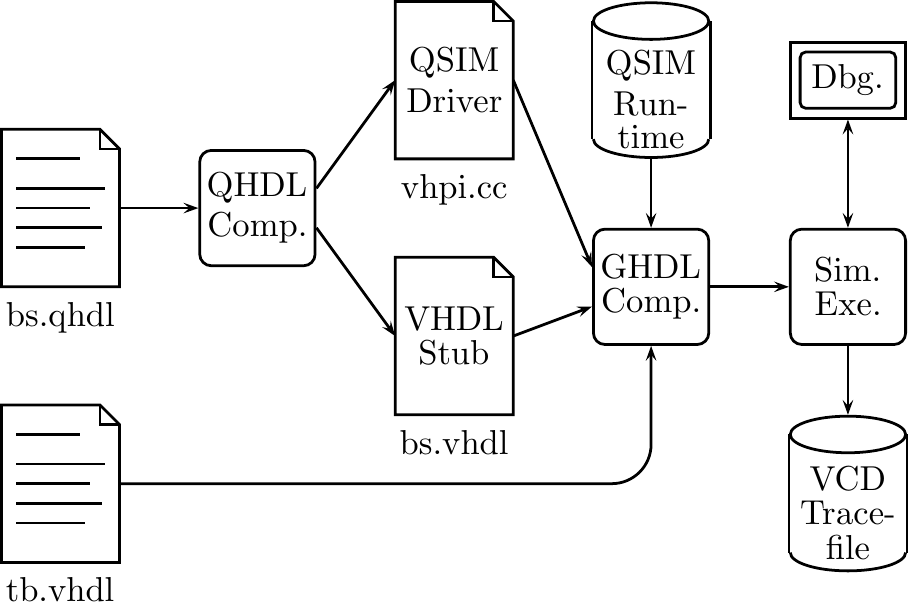}
  \caption{\label{fig:workflow}Typical workflow when using the QHDL/VHDL co-simulation approach.}
  \Description{Figure described in section 4.}
\end{figure}

\section{The QHDL Compiler and Simulator}
\label{sec:cosim}

As a proof-of-concept, we implemented a simple QHDL compiler and a
quantum simulator called QSIM that can be integrated
with a VHDL simulation using the open-source GHDL tools~\cite{ghdl}
to create a QHDL/VHDL co-simulation, see Figure~\ref{fig:workflow}.

The QHDL compiler analyses and elaborates the quantum circuit descriptions,
e.g. \texttt{bs.qhdl}, resulting in a flattened description of the circuit
consisting of built-in gates connected by signals, similar to a gate-level
netlist.
The compiler then infers the number of quantum bits necessary to describe
the state of the quantum circuit and the order of evaluation of the
different operations on this state to simulate the quantum circuit.
It then creates a VHDL wrapper, \texttt{bs.vhdl}, that interfaces to the VHDL
model, \texttt{tb.vhdl}, of the classical system, and a state evolution
function, \texttt{vhpi.cc}, that drives the quantum simulation.
The GHDL compiler is then used to analyze and elaborate the VHDL models,
compile the state evolution function and link everything together
with the QSIM runtime to create an executable simulator.

The QSIM state-vector simulation run-time library contains
the supporting infrastructure to simulate the quantum system and
the implementation of the built-in quantum gates.
The QSIM simulator can write the evolution of the system state to a
VCD trace file.
A built-in WEB-based debugger allows a user to single-step a running
QSIM simulation, see Figure~\ref{fig:belldbg}.
The QSIM simulator is aware of the simulation time of the VHDL model
of the classical circuit allowing to correlate its activity
with the quantum simulation.

\section{Use Case: Bell's Quantum Circuit}
\label{sec:bell}
\begin{figure}
  \includegraphics{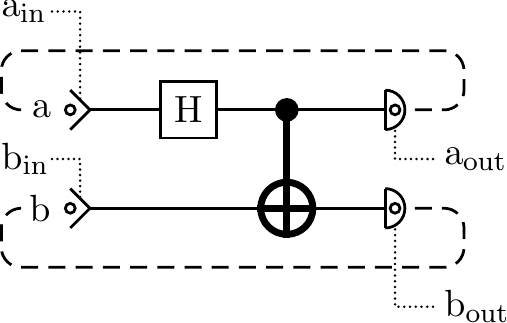}
  \caption{\label{fig:bellcirc}Bell's state circuit with additional feedback signals, dashed, and classical inputs and outputs, dotted.}
  \Description{A quantum circuit diagram to create a Bell's state. The circuit uses two qubits, labelled a and b, entering at the left edge of the diagram. Both qubits are simultaneously set-up from two classical inputs ain and bin. The top qubit, a, is after this fed into a Hadamard-gate and from this to the control input of a CNOT gate. The bottom qubit, b, is directly fed to the data input of the CNOT gate.  Both outputs from the CNOT gate are then measured with the top measurement result connected to a classical output aout and the bottom result connected to bout.  The qubit outputs from the two measurement opertions are fed back to the qubit inputs of the setup operations on the left of the diagram.}
\end{figure}

As an example of the usage of QHDL we have modelled a basic quantum circuit
to create a pair of entangled qubits, also known as Bell's pair.
As shown in Figure~\ref{fig:bellcirc} the circuit initializes 2 qubits, a and b,
from two classical inputs, $\mathrm{a_{in}}$ and $\mathrm{b_{in}}$,
applies two quantum gates, a Hadamard and a controlled-NOT gate, and then
measures both qubits producing the classical outputs,
$\mathrm{a_{out}}$ and $\mathrm{b_{out}}$.

The QHDL description of the circuit is shown in Figure~\ref{fig:bellqhdl}.
Lines 3 and 5 import the QHDL standard library.
In lines 7--13 the external interface is defined using an \emph{entity}
statement which also includes a clock signal for
synchronization with the VHDL simulation.
The \emph{architecture} body in lines 15--30 defines the implementation
of the circuit consisting of six component instantiations: Two setup
operations, lines 19--20 and 21--22; the Hadamard gate, line 23; the
controlled-NOT gate, lines 24--25; and two measurement operations,
lines 26--27 and 28--29.
Seven \texttt{qbit}-type signals, defined in lines 16--17,
connect the components, with two feedback-connections used to avoid unconnected
in- and outputs.

\begin{figure}
\begin{minipage}{80mm}
\begin{Verbatim}[fontsize=\scriptsize,numbers=left,frame=single,numbersep=4pt]
-- A pair of qbits that can be prepared in a Bell state

library qhdl;

use qhdl.std.all;

entity bellstate is
  port (
    clk: in bit;
    a_in, b_in: in bit;
    a_out, b_out: out bit
    );
end entity bellstate;

architecture quantum of bellstate is
  signal reg_a, reg_b, had_a, not_a, not_b,
      meas_a, meas_b: qbit;
begin
  setter_a: qset port map ( clk => clk, d => meas_a,
      q => reg_a, set => a_in );
  setter_b: qset port map ( clk => clk, d => meas_b,
      q => reg_b, set => b_in );
  hadamat_a: qhadamard port map ( d => reg_a, q=> had_a );
  entangle: qcnot port map ( c_in => had_a, c_out => not_a,
      d => reg_b, q => not_b );
  measure_a: qmeasure port map ( clk => clk, d => not_a,
      q => meas_a, result => a_out );
  measure_b: qmeasure port map ( clk => clk, d => not_b,
      q => meas_b, result => b_out );
end architecture quantum; -- bellstate
\end{Verbatim}
\end{minipage}
  \caption{\label{fig:bellqhdl}Bellstate circuit in QHDL.}
  \Description{The QHDL listing of the Bellstate circuit. 
  Line  1: -- A pair of qbits that can be prepared in a Bell state
  Line  2 blank
  Line  3: library qhdl;
  Line  4 blank
  Line  5: use qhdl.std.all;
  Line  6 blank
  Line  7: entity bellstate is
  Line  8:   port (
  Line  9:     clk: in bit;
  Line 10:     a\_in, b\_in: in bit;
  Line 11:     a\_out, b\_out: out bit
  Line 12:     );
  Line 13: end entity bellstate;
  Line 14 blank
  Line 15: architecture quantum of bellstate is
  Line 16:   signal reg\_a, reg\_b, had\_a, not\_a, not\_b,
  Line 17:      meas\_a, meas\_b: qbit;
  Line 18: begin
  Line 19:   setter\_a: qset port map ( clk => clk, d => meas\_a,
  Line 20:       q => reg\_a, set => a\_in );
  Line 21:   setter\_b: qset port map ( clk => clk, d => meas\_b,
  Line 22:       q => reg\_b, set => b\_in );
  Line 23:   hadamat\_a: qhadamard port map ( d => reg\_a, q=> had\_a );
  Line 24:   entangle: qcnot port map ( c\_in => had\_a, c\_out => not\_a,
  Line 25:       d => reg\_b, q => not\_b );
  Line 26:   measure\_a: qmeasure port map ( clk => clk, d => not\_a,
  Line 27:       q => meas\_a, result => a\_out );
  Line 28:   measure\_b: qmeasure port map ( clk => clk, d => not\_b,
  Line 29:       q => meas\_b, result => b\_out );
  Line 30: end architecture quantum; -- bellstate}
\end{figure}

The VHDL simulation of the non-quantum part of the example is shown
in Figure~\ref{fig:belltb} and is modelled as a typical HDL testbench,
entity \texttt{bellstate\_tb} that has no connections to the environment.
The behavior of the entity, and thereby of the VHDL simulation, is defined
by four processes:
The \texttt{clock} process, lines 10--13, drives the \texttt{clk} signal
with a 100\,MHz, 10\,ns, toggling clock with the first rising edge defined
at $t_1=5\,\mathrm{ns}$.
The \texttt{stimulus} process, lines 15--26, provides the two classic input
signals, set to constant '0', to the quantum part, and signals the
termination of the simulation after 101 clock-cycles.
The \texttt{tracer} process, lines 28--34, monitors the classic output signals
and prints the results to the console.
The \texttt{stats} process, lines 36--38, collects a histogram over all four
output states and reports those at the end of the simulation.
Finally, the quantum circuit is instantiated and connected to the classical
simulation in lines 40--42.

\begin{figure}
\begin{minipage}{80mm}
\begin{Verbatim}[fontsize=\scriptsize,numbers=left,frame=single]
-- A testbench for the bell-state circuit

entity bellstate_tb is
end entity bellstate_tb;

architecture testbench of bellstate_tb is
  signal clk, a_in, b_in, a_out, b_out: bit;
  signal done: boolean := false;
begin
  clock: process is
  begin
    -- Details omitted for brevity, create 100 MHz clock
  end process clock;

  stimulus: process is
    variable clk_cnt: natural := 0;
  begin
    -- Create always the same state
    a_in <= '0'; b_in <= '0';
    while clk_cnt < 101 loop
      wait until clk'event and clk='1';
      clk_cnt := clk_cnt + 1;
    end loop;
    done <= true;
    wait;
  end process stimulus;

  tracer: process (clk) is
  begin
    if clk'event and clk='1' then
      report "a_out=" & bit'image(a_out) &
          " b_out=" & bit'image(b_out);
    end if;
  end process tracer;

  stats: process (clk, done) is
    -- Details omitted for brevity, collect histogram
  end process stats;

  dut: entity work.bellstate
    port map ( clk => clk, a_in => a_in, b_in => b_in,
        a_out => a_out, b_out => b_out );

end architecture testbench; -- bellstate_tb
\end{Verbatim}
\end{minipage}
\caption{\label{fig:belltb}Bellstate testbench in VHDL with the details of the statistics gathering omitted.}
\Description[Abbreviated VHDL listing]{Line  1: -- A testbench for the bell-state circuit
Line  2 blank
Line  3: entity bellstate\_tb is
Line  4: end entity bellstate\_tb;
Line  5 blank
Line  6: architecture testbench of bellstate\_tb is
Line  7:   signal clk, a\_in, b\_in, a\_out, b\_out: bit;
Line  8:   signal done: boolean := false;
Line  9: begin
Line 10:  clock: process is
Line 11:  begin
Line 12:    -- Details omitted for brevity, create 100 MHz clock
Line 13:  end process clock;
Line 14 blank
Line 15:  stimulus: process is
Line 16:    variable clk\_cnt: natural := 0;
Line 17:  begin
Line 18:    -- Create always the same state
Line 19:    a\_in <= '0'; b\_in <= '0';
Line 20:    while clk\_cnt < 101 loop
Line 21:      wait until clk'event and clk='1';
Line 22:      clk\_cnt := clk\_cnt + 1;
Line 23:    end loop;
Line 24:    done <= true;
Line 25:    wait;
Line 26:  end process stimulus;
Line 27 blank
Line 28:  tracer: process (clk) is
Line 29:  begin
Line 30:    if clk'event and clk='1' then
Line 31:      report "a\_out=" \& bit'image(a\_out) \&
Line 32:          " b\_out=" \& bit'image(b\_out);
Line 33:    end if;
Line 34:  end process tracer;
Line 35 blank
Line 36:  stats: process (clk, done) is
Line 37:    -- Details omitted for brevity, collect histogram
Line 38:  end process stats;
Line 39 blank
Line 40:  dut: entity work.bellstate
Line 41:    port map ( clk => clk, a\_in => a\_in, b\_in => b\_in,
Line 42:        a\_out => a\_out, b\_out => b\_out );
Line 43 blank
Line 44: end architecture testbench; -- bellstate\_tb}
\end{figure}

\begin{figure} 
  \center
  \includegraphics[width=60mm]{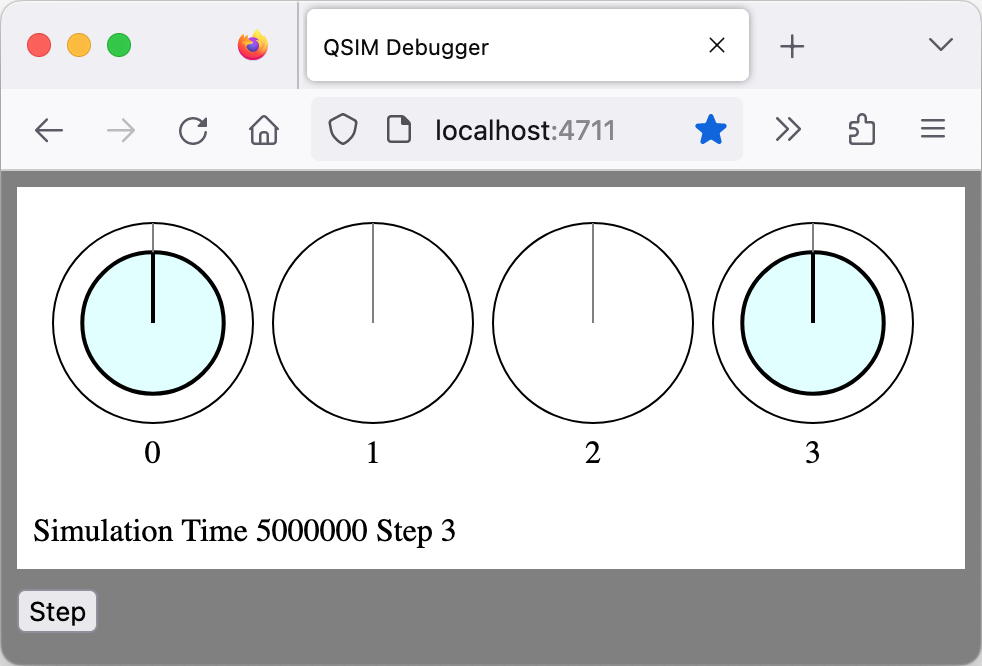}
  \caption{\label{fig:belldbg}Screenshot of the built-in QSIM debugger showing the superposition of entangled states 0 and 3 using the circular notation.}
  \Description{A screenshot of a web-browser window showing a page labelled QSIM debugger reachable at the URL localhost:4711. The webpage consists of two areas, a status display on top and a interaction area below it.  The interaction area has a single button labelled step. The status display shows four circles in a line with labels 0, 1, 2, 3 from left to right indicating the basic state below them. Both circles for state 0 and 3 show an inner circle in bold and colored with a radius of 70 per-cent of the outer circle to represent the magnitude of the probability amplitude. The phase of all four states is indicated with a line staight upwards from the center to the radius of each of the four outer circles representing a phase of zero. Below the circle display a separate line shows timing information: Simulation time 5000000 step 3.}
\end{figure}

The internal state, $\mathrm{| \psi \rangle}$, of the quantum simulation after
application of the controlled-NOT operation at the
first rising clock edge, step 3 of 0-5 at simulation time
$t_1=5\,000\,000\,\mathrm{fs}$, as visualised by
the built-in debugger is shown in Figure~\ref{fig:belldbg}.
The circular notation~\cite{johnston2019programming} shows both the magnitude and phase of the probability amplitude for each of the base states.

\section{Conclusions}
This paper presented QHDL, a low-level programming language akin to VHDL, for integrating FPGA and gate-based quantum computing programming approaches. A proof-of-concept QHDL/VHDL co-simulation environment was described. We envision that QHDL could be helpful to support process real-time, streaming data with low latency and high bandwidth requirements, perform advanced statistic gathering from repeated quantum computations, and implement closed-loop hybrid algorithms. While promising, the current work has a series of limitations, including better language support, improved debugging and tracing capabilities, the production of compilation reports, and optimization of the produced system.
Integration with pulse-level simulations would allow to back-annotate detailed timing information into the QHDL model.

\bibliographystyle{ACM-Reference-Format}
\bibliography{QHDL}
\end{document}